\documentclass[fleqn,12pt,twoside]{article}
\usepackage{espcrc1}

\usepackage{graphicx}
\usepackage[figuresright]{rotating}

\makeatletter
\def\simleq{\mathrel{\mathpalette\gl@align<}}
\def\simgeq{\mathrel{\mathpalette\gl@align>}}
\def\gl@align#1#2{\lower.6ex\vbox{\baselineskip\z@skip\lineskip\z@
     \ialign{$\m@th#1\hfill##\hfil$\crcr#2\crcr\sim\crcr}}}
\makeatother
\newcommand{\bec}[1]{\mbox{\boldmath $#1$}}
\newcommand{\fslash}[1]{\ooalign{\hfil/\hfil\crcr$#1$}}
\newcommand{\bra}{\langle}
\newcommand{\ket}{\rangle}
\newcommand{\braket}[1]{\bra #1 \ket}
\newcommand{\qq}{\braket{\bar{q}q}}
\renewcommand{\ss}{\braket{\bar{s}s}}

\newcommand{\qGq}{g\braket{\bar{q}\sigma_{\mu\nu}G_{\mu\nu} q}}
\newcommand{\sGs}{g\braket{\bar{s}\sigma_{\mu\nu}G_{\mu\nu} s}}
\newcommand{\gf}{\gamma_5}
\newcommand{\PV}{\gamma_5\fsl{p}}
\newcommand{\Tensor}{\gamma_5\sigma_{\mu\nu}q^\mu p^\nu}

\newcommand{\fsl}[1]{\fslash{#1}}

\newcommand{\pLS}{\pi\Lambda\Sigma}
\newcommand{\gpLS}{g_{\pi\Lambda\Sigma}}
\newcommand{\AmS}{{\protect\the\textfont2
  A\kern-.1667em\lower.5ex\hbox{M}\kern-.125emS}}

\title{The F/D Ratio and Meson-Baryon Couplings from QCD Sum Rules}

\author{Takumi Doi\address[Titech]{Dept. of Phys., 
			Tokyo Institute of Technology, 
			Meguro, Tokyo 152-8551, Japan}%
        \thanks{E-mail: doi@th.phys.titech.ac.jp},
        Hungchong Kim\address{Inst. of Phys. and Applied Phys., Yonsei University, 
		Seoul 120-749, Korea},
        Yoshihiko Kondo\address{Kokugakuin University, Higashi Shibuya, Tokyo 150-8440, Japan}
        and
        Makoto Oka\addressmark[Titech]
		}
       
\begin{document}

\maketitle

\begin{abstract}
Coupling constants of the pseudoscalar mesons 
to the octet baryons are computed in the QCD sum rule approach. 
The 
$\pi NN$, $\eta NN$, $\pi\Xi\Xi$, $\eta\Xi\Xi$, $\pi\Sigma\Sigma$, $\eta\Sigma\Sigma$ 
as well as $\pi\Lambda\Sigma$ couplings are studied. 
Determining the pertinent Dirac structure in the correlation function,
we analyze the couplings in the SU(3) limit.
We find the F/D ratio to be $\sim 0.6-0.8$ 
that is consistent with the SU(6) value. We also estimate
the SU(3) breaking effect using a projected correlation 
function method.
\end{abstract}

\section{INTRODUCTION}

Recent developments in hypernuclear physics require us knowledges 
of the interactions among strange hadrons,
 such as hyperons ($\Lambda$, $\Sigma$, \ldots) and kaons.   
Much efforts have been made to analyze experimental data 
to draw the consistent picture of the hyperon interactions. 
Yet the most fundamental quantities, 
i.e., the meson-baryon coupling constants, which are essential 
in constructing the meson exchange forces 
of baryons, have been treated as free parameters.  There the SU(3) 
invariance of the coupling constants is often 
employed without foundation.  Under these circumstances, 
it is highly desirable to calculate the meson-baryon 
coupling constants from the fundamental theory.

In this report, we present a QCD sum rule calculation of 
the coupling constants of the pseudoscalar mesons to 
the octet baryons.  The QCD sum rule has been successful in 
determining the masses and couplings of various hadrons 
in terms of  the QCD vacuum condensates and the other fundamental 
constants.  In applying the sum rule technique,  
several technical points are carefully considered first.  
The soft pion (and other pseudoscalar mesons)  is often used 
in deriving the sum rule for the coupling constants.  
We stress that the finite meson momentum should be considered 
in order to obtain reliable sum rule.  We need to choose the Dirac 
structure of the correlation function carefully so that the sum rule 
is practically workable.  It is also necessary to check dependence 
on the choice of the interpolating field operator for the baryon.  
The last point is especially important in deriving the F/D ratio 
for the coupling constants.

In section 2, we present the formulation for the sum rule of 
baryon-diagonal coupling constants and the results of the Borel 
sum rule analyses.  We derive the F/D ratio in the analysis in the 
SU(3) limit.  
In section 3, we derive the sum rule for the $\pi\Lambda\Sigma$ 
coupling constant as an example of the baryon-off-diagonal case, 
where we use the projection method to define the coupling constant 
unambiguously.

\section{SUM RULE FOR THE MESON-BARYON COUPLING CONSTANTS}

We analyze baryon-diagonal meson-baryon couplings ${\cal M}BB$ for 
meson ${\cal M}=\pi,\eta$ and baryon $B=N,\Xi,\Sigma$.
We construct QCD sum rules 
from the two-point correlation 
function with an external meson field
\begin{eqnarray}
\label{eq:corr}
\Pi(q,p)
= i\int d^4x e^{iqx} \bra 0|{\rm T}[J_B(x)\bar{J}_B(0)] |{\cal M}(p)\ket,
\end{eqnarray}
where $J_B$ is general baryon current without derivative.
The explicit formula for $J_N$ is 
%
%
$
J_N(t) = 2\epsilon _{abc}
[\ (u_a^T C d_b)\gamma_5 u_c
+ t\ (u_a^T C\gamma_5 d_b)u_c\ ],
%
$
%
%
and $J_\Xi$ and $J_\Sigma$ are obtained via
SU(3) rotations of $J_N$.
We note general currents have free mixing parameter $t$, which is 
essential to check the reliability of sum rules later.
It is known that if one takes 
the soft-meson limit $p\rightarrow 0$, 
the sum rule from 
Eq.~(\ref{eq:corr}) becomes equivalent to the chiral-rotated 
nucleon mass sum rule via the soft-meson theorem~\cite{BK},
and reduces to the Goldberger-Treiman
relation with $g_A=1$~\cite{SH}.
Therefore, it is necessary to go beyond the soft-meson limit
and determine the coupling constants independently of
mass sum rule.

When going beyond the soft-meson limit, 
we have three distinct Dirac structures
($i\gamma_5\fslash{p}$, $i\gamma_5$ and 
$\gamma_5\sigma_{\mu\nu}q^\mu p^\nu$)
in the correlation function~\cite{KD1,KD2,DK1}  as
\begin{eqnarray}
\Pi =
  i\gamma _5 \fslash{p}\ \Pi _{\rm PV}
+ i\gamma _5\ \Pi _{\rm PS} 
+ \gamma _5 \sigma ^{\mu\nu} q_\mu p_\nu\ \Pi _{\rm T}.
\end{eqnarray}
From each Dirac structure, we obtain sum rules for
the coupling constant $g_{{\cal M}B}$ as
\begin{eqnarray}
\label{eq:sumrule}
g_{{\cal M}B}\lambda_B^2(t)\left[\ 1+ A_{{\cal M}B} (t) M^2 \ \right] 
\ = \ 
f^{\mbox{\tiny OPE}}_{\mbox{\tiny\it ${\cal M}$B}}(M^2;t)\ ,
\end{eqnarray}
where $M$ denotes the Borel mass,
$\lambda_B(t)$ overlap constant between the current and the corresponding baryon,
$A_{{\cal M}B}$ unknown parameter.
We calculate OPE term
$f^{\mbox{\tiny OPE}}_{\mbox{\tiny\it ${\cal M}$B}}(M^2;t)$
up to dimension 8 at ${\cal O}(p^2)$ ($i\gf$ structure),
and up to dimension 7 at ${\cal O}(p)$ ($i\PV$ and $\Tensor$ structures)~\cite{DK1}.
Considering $m_q \sim {\cal O}(p^2)$ indicated by 
the Gell-Mann--Oakes--Renner relation,
we include ${\cal O}(m_q)$ terms only in $i\gf$ structure.
By linearly fitting r.h.s. of Eq.~(\ref{eq:sumrule})
against $M^2$,
we determine 
$\left[ g_{{\cal M}B}\lambda_B^2(t) \right]_{\mbox{\scriptsize fitted}}$.

\subsection{Pertinent Dirac structure for coupling sum rules}

In principle, sum rules from any Dirac structure
$i\gamma_5\fslash{p}$, $i\gamma_5$ and 
$\gamma_5\sigma_{\mu\nu}q^\mu p^\nu$
can be used
to determine the couplings. In practice, however, 
different prediction is derived because 
reliability of the sum rule varies depending on the Dirac structure.
Therefore one must carefully choose pertinent Dirac structure.




First, we point out 
that the sum rules from $i\PV$ structure is highly
sensitive to the continuum threshold $S_0$ and thus is not reliable.
On the other hand, the sum rules from $i\gf$ and $\Tensor$ structure
are insensitive to $S_0$ and more appropriate
than those from $i\PV$ structure~\cite{KD1,KD2,DK1}.


%
%

Second, we check the dependence on the choice of the baryon current.
The $t$ dependence in Eq.~(\ref{eq:sumrule}) gives a new constraint.
In the SU(3) limit (i.e. $\lambda_N=\lambda_\Xi=\lambda_\Sigma$),
$g_{{\cal M}B}\lambda_{B}^2(t)$
has a common $t$ dependence and thus the ratios of different couplings
are $t$ independent.
We find that the sum rules from $\Tensor$ structure satisfy the 
above constraint well, while the sum rules from $i\gf$ structure 
do not.
Therefore, we conclude that $\Tensor$ structure 
is the most pertinent Dirac structure for coupling sum rules.

\subsection{Determination of the $F/D$ ratio}

In the SU(3) limit, 
we confirm that our sum rules satisfy the SU(3) relation
between couplings.
We first determine 
$\left[ g_{{\cal M}B}\lambda_B^2(t) \right]_{\mbox{\scriptsize fitted}}$
by linear fitting Eq.~(\ref{eq:sumrule})
for every ${\cal M}=\pi,\eta$ and $B=N,\Xi,\Sigma$.
We calculate the ratio between them and convert it into the $F/D$ ratio.
Note that ambiguity of $\lambda_B$ cancels 
because $\lambda_N=\lambda_\Xi=\lambda_\Sigma$ in the SU(3) limit.

\begin{figure}[bth]
\begin{minipage}[t]{78mm}
\includegraphics[scale=0.3]{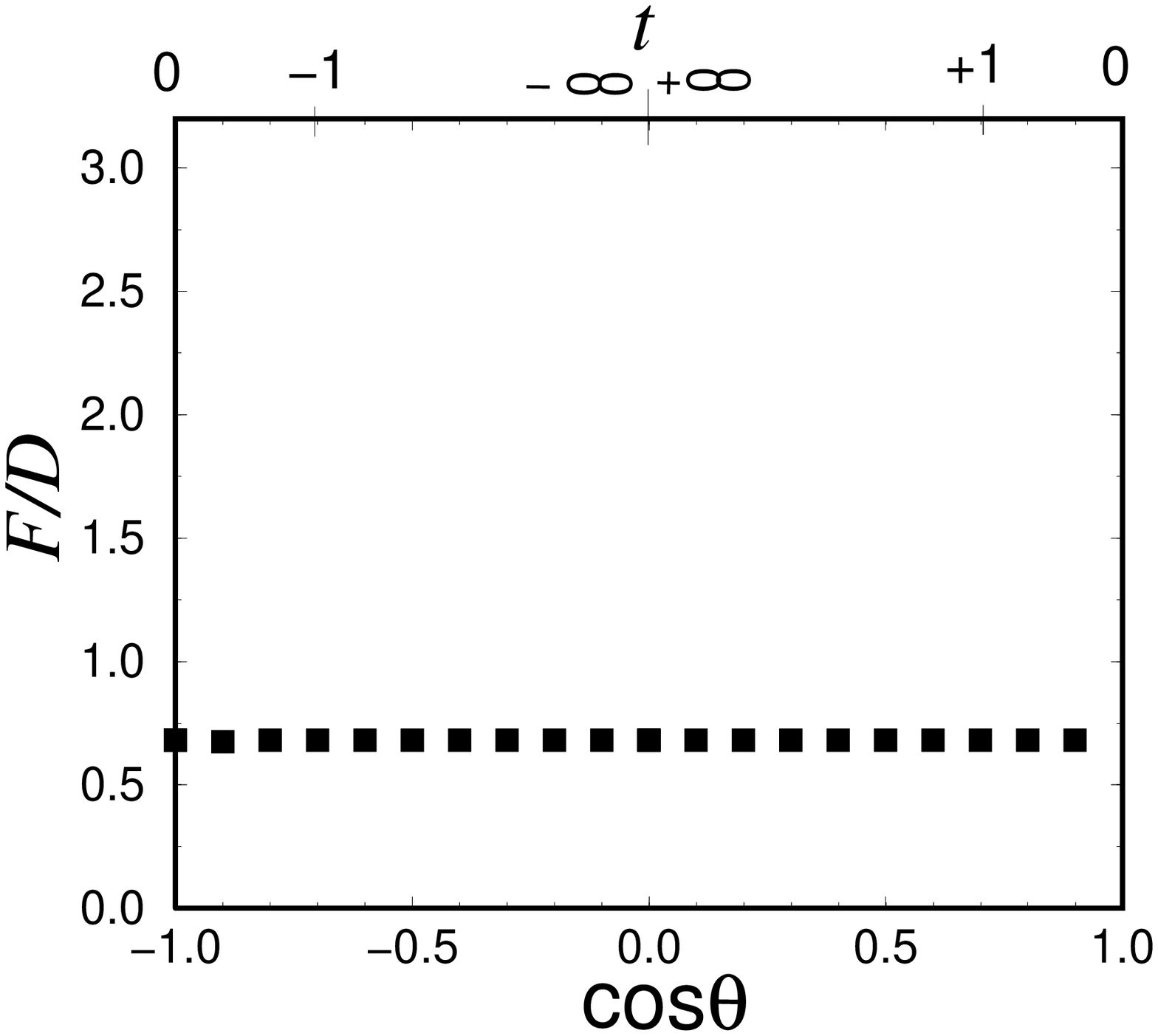}
\vspace*{-10mm}
\caption{The $F/D$ ratio from $\gamma_5\sigma_{\mu\nu}q^\mu p^\nu$ 
   structure against $\cos\theta$.
   Corresponding $t$ is also shown at the top of the figure.
}
\label{fig:FD_T-L}
\end{minipage}
\hspace{\fill}
\begin{minipage}[t]{77mm}
\includegraphics[scale=0.3]{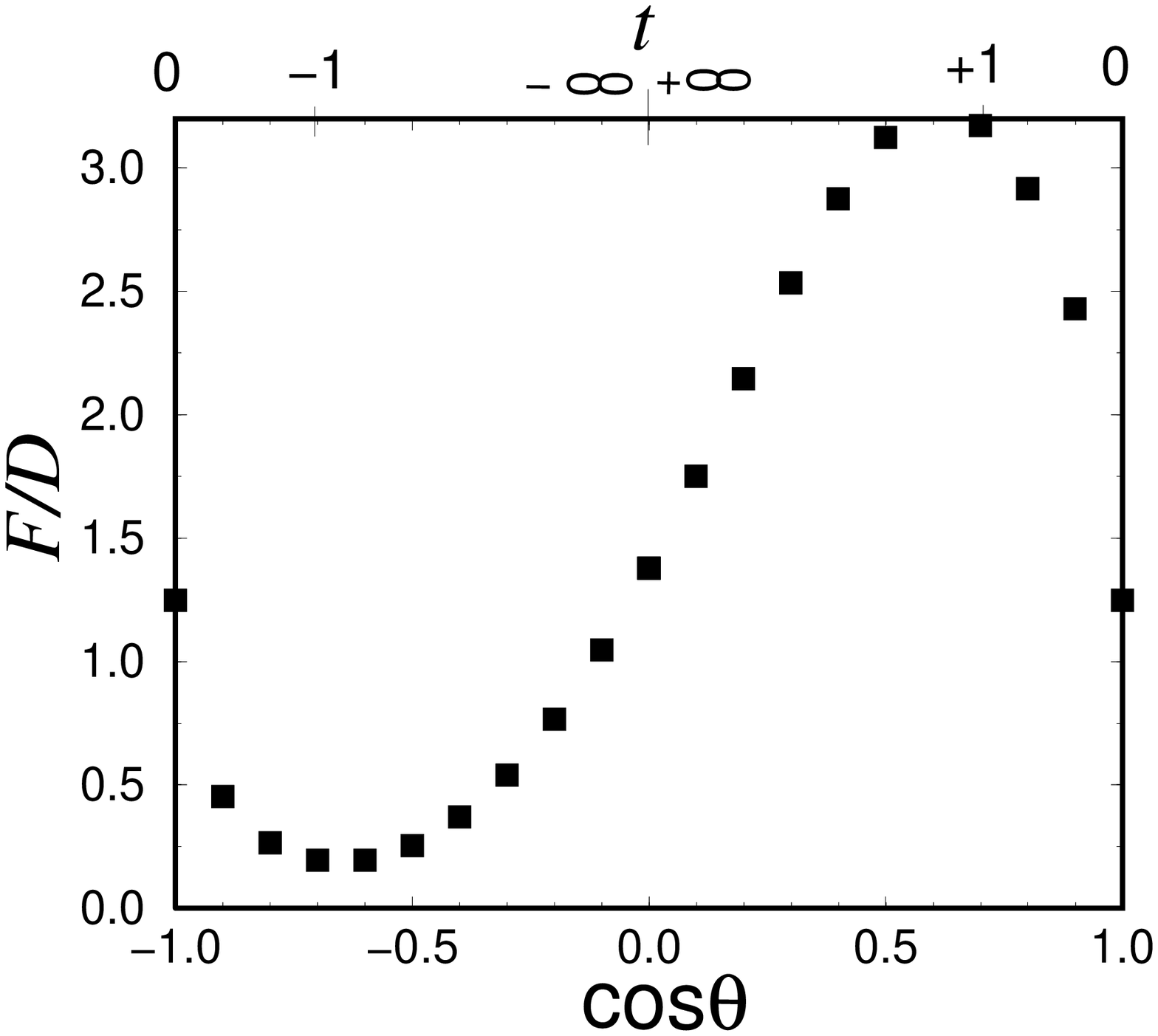}
\vspace*{-10mm}
\caption{The $F/D$ ratio from $i\gamma_5$ structure 
   against $\cos\theta$. 
}
\label{fig:FD_PS}
\end{minipage}
\end{figure}

In figure~\ref{fig:FD_T-L}, we show the $F/D$ ratio against the choice 
of baryon current. Here, we introduce a new parameter $\theta$,
which is defined as $\tan\theta = t$.
As discussed before, we confirm that the $F/D$ ratio 
is insensitive to both of the continuum threshold $S_0$ and
the parameter $\theta$ in the realistic region 
($-0.78 \protect\simleq \cos\theta \protect\simleq 0.61$)~\cite{DK1}.
Therefore, we can make a reliable estimate, $F/D = 0.6-0.8$.
This range includes the value from the SU(6) quark
model ($F/D = 2/3$),
and is slightly higher than that extracted from 
semi-leptonic decay rates of 
hyperons ($F/D \sim 0.57$)~\cite{ratcliffe}.
Just as a comparison, we also show 
the $F/D$ ratio obtained from $i\gf$ structure in figure~\ref{fig:FD_PS}.
We find that the prediction highly depends on the choice of $\theta$,
which prevents reliable estimate for the $F/D$ ratio.

\section{THE $\pLS$ COUPLING AND THE SU(3) BREAKING EFFECT}

We analyze the baryon-off-diagonal coupling $\pi\Lambda\Sigma$,
and its SU(3) breaking effect using a projection method~\cite{Kondo}.
The advantages of this method are the followings: 
(1)
The coupling constant is determined 
without referring to {\it a priori} 
effective Lagrangian,
(2) baryon mass difference is taken into account properly, and
(3) the single-pole terms and the continuum terms
are well-defined in the correlation function.
In this scheme, we calculate the projected correlation function
defined as
\begin{eqnarray}
\Pi_+ (q,p) &\equiv& \bar{u}_\Sigma (\bec{q})\ \gamma_0 \Pi (q,p)
\gamma_0\ u_\Lambda (\bec{q}-\bec{p}) ,
\end{eqnarray}
where $\Pi(q,p)$ is given by Eq.~(\ref{eq:corr}) and $u(\bec{q})$ 
is the Dirac spinor for the plane wave with momentum $q$.
In the phenomenological side, we obtain $\gpLS$ (for $\pi^0\Lambda\Sigma^0$) as 
\begin{eqnarray}
\lefteqn{
-\lambda_\Lambda \lambda_\Sigma\cdot g_{\pi\Lambda\Sigma}\cdot \bar{u}_{\Sigma}(\bec{q})
 i\gamma_5 u_{\Lambda}(\bec{q}-\bec{p})
} \nonumber \\ 
&=&
\bar{u}_{\Sigma} (\bec{q}) 
(\fslash{q} - M_\Sigma ) \Pi (q,p) (\fslash{q}-\fslash{p}
- M_\Lambda ) 
u_\Lambda (\bec{q}-\bec{p})
\Bigr|_{q^2=M_\Sigma^2, (q-p)^2 = M_\Lambda^2} \\
&=& (q_0-M_\Sigma) (q_0 - E_p) \Pi_+ (q) ,
\end{eqnarray}
where $E_p \equiv \sqrt{M_\Lambda^2 + \bec{p}^2} + \sqrt{m_\pi^2 + \bec{p}^2}$.
We use the dispersion relation with respect to $q_0$ for 
$\Pi_+^{\rm odd} \equiv \frac{1}{2q_0}\left[ \Pi_+(q_0,\bec{q}) - \Pi_+(-q_0,\bec{q}) \right]$
to improve the sum rule. We note that though all Dirac structures in $\Pi(q,p)$ are mixed 
in $\Pi_+(q,p)$, $\Pi_+^{\rm odd}(q,p)$ enhances the contribution from $\Tensor$ structure
in $\Pi(q,p)$.
We calculate the OPE up to dimension 7, including terms proportional to the quark mass.
We use the baryon mass sum rules to eliminate $\lambda_B$, and 
extract $\gpLS$ by operating $(1- M^2 \frac{\partial}{\partial M^2})$.

\begin{figure}[bth]
\begin{center}
\includegraphics[scale=0.3]{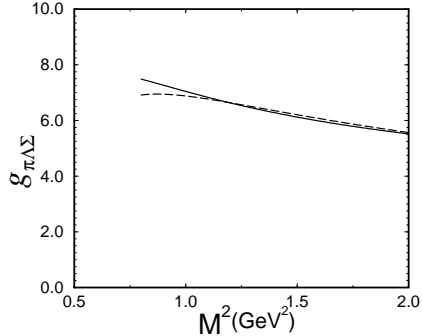}
\vspace*{-10mm}
\caption{$\gpLS$ against the Borel mass $M^2$.
The solid line denotes the SU(3) limit, the dashed line
beyond SU(3) limit.
}
\label{fig:pi-lambda-sigma}
\end{center}
\end{figure}

In figure~\ref{fig:pi-lambda-sigma}, we show preliminary Borel plot from 
our result without continuum effect.
We find that the Borel stability is obtained
and we estimate $\gpLS \simeq 6-7 $.
Further, we include the SU(3) breaking effects via
$m_s\neq m_q$, $\ss \neq \qq$, $\sGs \neq \qGq$, $M_B (B=\Lambda,\Sigma) \neq M_N$,
where $q$ denotes u, d-quark.
We show the Borel plot in figure~\ref{fig:pi-lambda-sigma}.
From this figure, we see that the SU(3) breaking effect is small.
More conclusive study including the continuum effects 
is in progress now~\cite{DKoO}.

\end{document}